\documentclass[conference]{IEEEtran}
\usepackage{amssymb}
\usepackage{graphicx}
\usepackage{multirow}

\begin{document}
\title{A Case-Study on Teaching Undergraduate-Level Software Engineering Course Using Inverted-Classroom, Large-Group, Real-Client and Studio-Based Instruction Model}

\author{
    \IEEEauthorblockN{Ashish Sureka, Monika Gupta, Dipto Sarkar, Vidushi Chaudhary} \\
    \IEEEauthorblockA{Indraprastha Institute of Information Technology, Delhi (IIITD) \\ New Delhi (India) \\
    \{ashish, monikag, diptos, vidushi1116\}@iiitd.ac.in}}

\maketitle
\IEEEpeerreviewmaketitle
\begin{abstract}
We present a case-study on teaching an undergraduate level course on Software Engineering (second year and fifth semester of bachelors program in Computer Science) at a State University (New Delhi, India) using a novel teaching instruction model. Our approach has four main elements: inverted or flipped classroom, studio-based learning, real-client projects and deployment, large team and peer evaluation. We present our motivation and approach, challenges encountered, pedagogical benefits, findings (both positive and negative) and recommendations. Our motivation was to teach Software Engineering using an active learning (significantly increasing the engagement and collaboration with the Instructor and other students in the class), team-work, balance between theory and practice, imparting both technical and managerial skills encountered in real-world and problem-based learning (through an intensive semester-long project). We conduct a detailed survey (anonymous, optional and online) and present the results of student responses. Survey results reveal that for nearly every students (class size: 89) the instruction model was new, interesting and had a positive impact on the motivation in addition to meeting the learning outcome of the course.  
\end{abstract}
\begin{keywords}
Software Engineering Education, Project-Based Learning, Studio-Based Learning, Inverted Classroom Instruction Model, Teaching Methodology
\end{keywords}

\begin{figure*}[t]
\centering
\vspace{-2.0cm} 
\includegraphics[scale=0.55, angle=0]{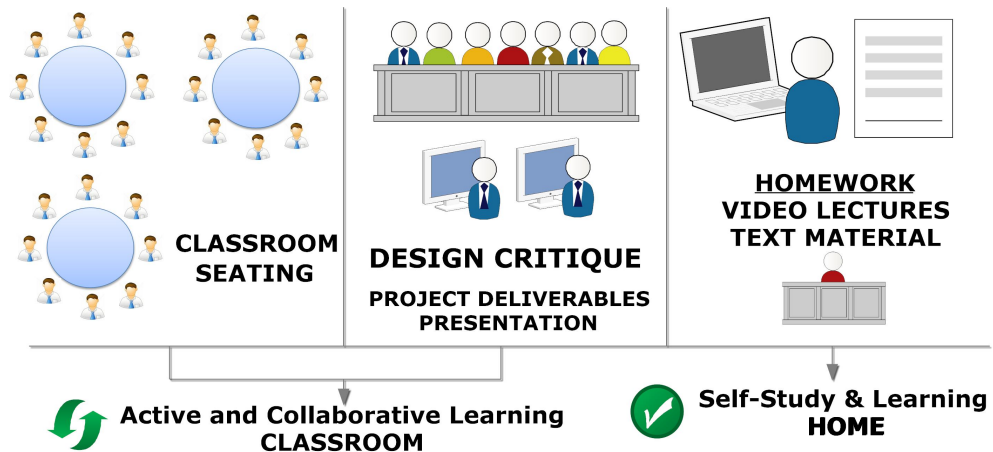}
\vspace{-2.7cm} 
\caption{Broad Framework of the One Semester $4$ Credit UG Level SE Course emphasizing active \& collaborative learning, real-client driven project-based, development of self-study and team skills}\label{fig:framework}
\end{figure*}

\begin{table*}[t]
\caption{Survey Results: Large Team Setting and Peer Evaluation in Software Engineering Course Project\label{tab:largegroup}}
\begin{center}
\begin{tabular}{|p{16cm}|c|}
\hline 
\multicolumn{2}{|p{18cm}|}{\textbf{\small Q1: Is this your first experience in developing a software-based solution in an 8 member team setting for a duration of 4-5 months?}} \\ \hline
\footnotesize Being part of an 8 member software project team for a substantial duration (4-5 months) was my first experience & \footnotesize 97.2\% \\ \\
\footnotesize I have previous experience of working in large software project team but this is my first large-team experience for such a size, duration and complexity & \footnotesize 2.8\% \\
\\ \footnotesize I have been involved in software projects consisting of more than 8 members and projects which were larger and more complex than my SE course project & \footnotesize 0.0\% \\ \hline

\multicolumn{2}{|p{18cm}|}{\textbf{\small Q2: What is your experience on the management overhead incurred due to managing a large group (8 members) in contrast to managing a small group (less than or equal to 5 members)?}} \\ \hline
\footnotesize Encountered significant management overhead and difficulties due to managing a large team & \footnotesize 61.1\% \\ \\
\footnotesize Little bit (modest) of overhead and difficulties as we reduced the overhead by defining processes and protocols and used project management tools for co-ordination & \footnotesize 30.6\% \\
\\ \footnotesize Did not encounter any overhead or major issues and we were able to manage very effectively using processes and tools & \footnotesize 8.3\% \\ \hline

\multicolumn{2}{|p{18cm}|}{\textbf{\small Q3: Do you think working in a large group exposed you to project management challenges (not present in small group setting) which were new to you?}} \\ \hline
\footnotesize I encountered several problems and challenges by working in a large team which were new to me and which I did not encounter my previous software projects & \footnotesize 72.2\% \\ \\
\footnotesize Few minor differences (in terms of new experiences and challenges) but not major difference from my past experiences & \footnotesize 25.0\% \\
\\ \footnotesize The management problems and challenges that I faced by working in a large group were not new to me & \footnotesize 2.8\% \\ \hline

\multicolumn{2}{|p{18cm}|}{\textbf{\small Q4: Did you think the anonymous peer-evaluation method to assess individual contribution in a large-group setting is needed to bring objectivity and fairness?}} \\ \hline
\footnotesize I strongly support anonymous peer-evaluation to evaluate individual contribution in large-group setting & \footnotesize 13.9\% \\ \\
\footnotesize I support peer-evaluation in large-group setting but I feel that it is prone to manipulation & \footnotesize 77.8\% \\
\\ \footnotesize I believe everyone in the group should be awarded same marks and do not support peer-evaluation & \footnotesize 8.3\% \\ \hline

\multicolumn{2}{|p{18cm}|}{\textbf{\small Q5: Do you think working in a large-group setting helped you better appreciate and understand concepts on version control system, project wiki, issue tracking systems and project management tools such as GANTT charts?}} \\ \hline
\footnotesize Working in large-group setting was a good mechanism to provide exposure to such tools & \footnotesize 52.8\% \\ \\
\\ \footnotesize Working in large group did not make any difference and such tools can be learnt equally well in a small group setting also & \footnotesize 47.2\% \\ \hline
\end{tabular}
\end{center}
\end{table*}

\section{Learning Objectives and Course Framework}
Software Engineering (SE) education is an area that has attracted several researchers' attention. Teaching SE at undergraduate level poses several challenges to the instructor and educators of SE have experimented (and presented their work through research papers and experience reports) with different instruction models depending on their motivation, learning objectives and the context. In this paper, we (Instructor and Teaching Assistants) present an experience report on teaching SE at undergraduate level using a novel instruction model. Software Engineering (CSE $300$) is a $4$ credit core course (two classes of $1.5$ hours each in one week with a total of $26$ classes over $4$ months) offered during the third year ($5^{th}$ semester) of Bachelor of Technology (abbreviated as B.Tech) in Computer Science and Engineering (CSE) program at Indraprastha Institute of Information Technology (IIIT Delhi, a state university in India). This experience report is based on SE course that we taught during Monsoon $2012$ semester (August $2012$ - December $2012$) with a class size of $89$ students.  

Figure \ref{fig:framework} is the broad framework of our course. Figure \ref{fig:framework} shows the classroom seating arrangement (motivated by Studio Based Teaching and Learning Model) facilitating group-work, active and collaborative learning. The course consists of several intensive design critique and project (real-client and large team-based) presentation sessions motivated by bringing active learning in classroom. One of the main elements of the course is self-study and self-learning using online video lectures and text material (through an inverted classroom setting) complemented or reinforced by discussion and clarification in classroom or teaching team office hours. The four main elements of our instruction model are: large-group and peer evaluation, real-client based project and deployment, studio-based learning and inverted classroom. Each of the four elements is discussed in detail in the following sections. Table \ref{tab:largegroup}, \ref{tab:realclient}, \ref{tab:studio} and \ref{tab:flipped} displays the results of the survey (online, optional and anonymous) conducted by us on all the four mains aspects of our instruction model.   
\begin{figure*}[t]
\centering
\includegraphics[scale=0.40, angle=0]{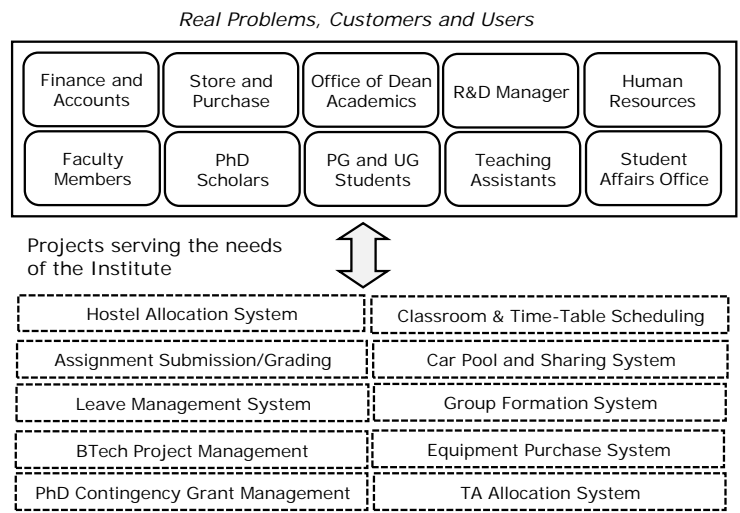}
\vspace{-1.1cm} 
\caption{List of projects developed by students of SE course and real users and customers with whom the students interacted (projects aimed at serving the needs of the Institute)}\label{fig:realclientp}
\end{figure*}
\section{Large Group and Peer Evaluation}
Team work and collaboration is integral to software development and engineering.  One of the learning objectives of the course is to learn team skills, study some of the best practices of team work in the context of software development and apply it in the semester-long team-based course project and experience both the benefits and challenges of working in a team. We created a team of 8-9 members each (a relatively large team in the context of a one-semester undergraduate-level course) as our aim was to motivate and demonstrate the need of good communication, cooperation, time-management, interdependence and reliance on other members, adjustment and negotiation.  Coppit et al. present their experience on teaching a one-semester software engineering course (at The College of William and Mary) in which $20$ to $30$ member student team develop a moderately sized ($15$ KLOC) software system and demonstrate the pedagogical benefits of large team \cite{coppit2006}. Student team projects are a hallmark of the undergraduate Software Engineering program at the Rochester Institute of Technology \cite{Reichlmayr2005}. Hogan et al. based on their work at Queensland University of Technology mention that having a tutor who facilitates team-work skills and someone who provides advice and guidance on team skills can improves team work skills \cite{Hogan2005}. Stein et al. present their insights on large vs. small group projects in capstone and software engineering courses \cite{Stein2002}. We asked $89$ students to create $8$ member teams on their own (self-organization rather than instructor created teams). However, we observe case-studies (two-semester software engineering course at the University of Texas at El Paso) in software engineering education in which instructor creates teams based on experience, skills and personalities \cite{Gates2000}.  
 
We instructed students to use Wiki Tab to create Wiki pages for their projects and use Google Project Hosting Wiki feature to document their project. Wiki based web-tools are popular for creating and sharing project documents enabling collaboration and team-work. All the members of the project team had permissions to add, modify and delete their respective project related documents and were helpful in knowledge management and project communication. We covered GANTT Charts during initial lectures and student teams created GANTT chart based project schedules. We believe that the application of GANTT chart for work breakdown, planning and scheduling was required due to a large team setting.    

Hayes et al. discuss several grading criteria (fair, consistent, reflect educational objectives, provide feedback, encourage students, no grade inflation, easy on grader) and grading schemes (such as all same grade, each reports own efforts, each evaluates self and others) as well combination of schemes \cite{Clark2005}\cite{Hayes2003}\cite{Wilkins2000}. We adopt a peer-evaluation based system to measuring individual contributions in a team-based project. Our project grading scheme was based on quality of outcome and peer. Clark et al. discuss guidelines (answering questions by peers on a pre-defined scale) for peer evaluation such as attendance at team meetings, contribution level in terms of giving ideas and assigned tasks and the performance of an individual in meeting deadlines for the assigned tasks. We instructed students to submit peer evaluation (distribute $70$ points across $7$ peers with the condition that the maximum marks which can be assigned to an individual is $20$) after every project deliverable ($5$ project deliverables). We experienced disputes related to peer-evaluation in some groups and few students reported problems related to collusion, intentionally giving low or high marks to peers, lack of contribution (free-riding) by project team members. Table \ref{tab:largegroup} present results of our survey conducted on large team setting and peer evaluation. Majority of the students ($77.8\%$) supported peer evaluation in large group setting but feel that it is prone to manipulation. Survey results reveal that majority of the students ($72.2\%$) encountered several problems and challenges by working in a large team which were new to them and which they did not encounter in their previous software projects.  

\section{Real Client Based Project and Deployment}
\begin{table*}[t]
\caption{Survey Results: Real Client Based Project and Deployment\label{tab:realclient}}
\begin{center}
\begin{tabular}{|p{16cm}|c|}
\hline 
\multicolumn{2}{|p{18cm}|}{\textbf{\small Q1: Is this your first hands-on experience in developing a software-based solution for real-client requirements and with the intent of deployment?}} \\ \hline
\footnotesize This is the first time I am involved in developing a software-based solution for a real-client & \footnotesize 81.5\% \\ \\
\footnotesize This is not my first real-client project but first of such a size, duration and complexity & \footnotesize 16.7\% \\
\\ \footnotesize This is not my first real-client project and I have been involved in software-based real-client projects which were larger and more complex than my SE course project & \footnotesize 1.9\% \\ \hline

\multicolumn{2}{|p{18cm}|}{\textbf{\small Q2: Was developing a solution aimed at serving the needs of your University (and solving everyday problems encountered by students) a motivator?}} \\ \hline
\footnotesize Increased my motivation substantially as my project is aimed at serving the needs of my University & \footnotesize 68.5\% \\
\\ \footnotesize Solving a problem relevant to my university did not have any additional major influence on my motivation & \footnotesize 31.5\% \\ \hline

\multicolumn{2}{|p{18cm}|}{\textbf{\small Q3: Did you encounter non-technical problems such as non-availability of the client when needed or lack of interest and time from the client resulting in loss of productivity for you?}} \\ \hline
\footnotesize Major problems & \footnotesize 22.2\% \\
\footnotesize Minor problems & \footnotesize 70.4\% \\
\footnotesize No problems & \footnotesize 7.4\% \\ \hline

\multicolumn{2}{|p{18cm}|}{\textbf{\small Q4: Do you think that the time-span of one-semester is enough from requirement analysis to user-acceptance testing and deployment for a real-client based course-project (in a 4 credit course, 8 member team)?}} \\ \hline
\footnotesize More than sufficient & \footnotesize 5.6\% \\
\footnotesize Just about sufficient or nearly sufficient & \footnotesize 35.2\% \\
\footnotesize Not sufficient & \footnotesize 59.3\% \\ \hline

\multicolumn{2}{|p{18cm}|}{\textbf{\small Q5: Do you think that developing a solution for a real-client played an important role and was helpful in learning managerial aspects, project and team management skills, communication and coordination with stakeholders?}} \\ \hline
\footnotesize Not helpful & \footnotesize 1.9\% \\
\footnotesize Little helpful but not much helpful & \footnotesize 37.0\% \
\\ \footnotesize Very helpful & \footnotesize 61.1\% \\ \hline
\end{tabular}
\end{center}
\end{table*}
\begin{table*}[t]
\caption{Survey Results: Studio-Based Teaching and Instruction Model\label{tab:studio}}
\begin{center}
\begin{tabular}{|p{16cm}|c|}
\hline 
\multicolumn{2}{|p{18cm}|}{\textbf{\small Q1: Is this your first experience of studio-based learning methodology (iterative design solution, design critiques and review sessions, separate work-space for collaborative working in classroom)?}} \\ \hline
\footnotesize This is my first experience with studio-based teaching method and instruction model & \footnotesize 100\% \\
\\ \footnotesize I have experienced studio-based teaching method and instruction model in past & \footnotesize 0.0\% \\ \hline

\multicolumn{2}{|p{18cm}|}{\textbf{\small Q2: What is your experience with the design critique and review sessions (presenting your artifacts on business process model, horizontal prototype, database design and functional testing)?}} \\ \hline
\footnotesize The design critique and review sessions in studio-based learning method were quite in-depth and critical in comparison to presentation and review sessions in traditional (lecture-based) courses and helped in increasing my understanding of the concepts & \footnotesize 73.5\% \\
\\ \footnotesize The depth and critical analysis of my presentation in design critique and review session in studio-based learning is equivalent or less than similar sessions in traditional (lecture-based) courses & \footnotesize 26.5\% \\ \hline

\multicolumn{2}{|p{18cm}|}{\textbf{\small Q3: What is your experience with the studio-based seating arrangement (chairs arranged in a circle and activity-based learning) in terms of teamwork, collaboration, degree of interaction and communication with group members)?}} \\ \hline
\footnotesize The studio-based arrangement increases teamwork, collaboration, degree of interaction and communication with group members & \footnotesize 47.1\% \\
\\ \footnotesize The studio-based arrangement does not have any noticeable increase in terms of teamwork, communication and collaboration with group members in comparison to courses (having tam-based project component) following traditional lecture-based instruction model & \footnotesize 52.9\% \\ \hline

\multicolumn{2}{|p{18cm}|}{\textbf{\small Q4: What is your experience with the studio-based seating arrangement (chairs arranged in a circle and activity-based class) in terms of motivation, excitement and engagement level?}} \\ \hline
\footnotesize The studio-based teaching method and instruction model increased my motivation, excitement and overall engagement level. & \footnotesize 58.8\% \\
\\ \footnotesize The studio-based teaching method and instruction model does not have any noticeable effect on my motivation, excitement and overall engagement level & \footnotesize 41.2\% \\ \hline
\end{tabular}
\end{center}
\end{table*}
\begin{table*}[t]
\caption{Survey Results: Flipped-Classroom Based Teaching Methodology Model\label{tab:flipped}}
\begin{center}
\begin{tabular}{|p{16cm}|c|}
\hline 
\multicolumn{2}{|p{18cm}|}{\textbf{\small Q1: Is this your first experience with flipped-classroom based teaching methodology (also called as reverse instruction – doing practical work, hands-on problem-solving exercise, team-based collaborative work in the classroom)?}} \\ \hline
\footnotesize This is my first experience with flipped-classroom based teaching method and instruction model & \footnotesize 95.0\% \\
\\ \footnotesize I have experienced flipped-classroom based teaching method and instruction model in past & \footnotesize 5.0\% \\ \hline

\multicolumn{2}{|p{18cm}|}{\textbf{\small Q2: What is your experience of flipped-classroom based teaching method in-terms of one-on-one interaction with the Instructor?}} \\ \hline
\footnotesize Since instructor is not lecturing - I have more opportunity of a one-on-one interaction with the Instructor in contrast to the lecture-based model. Having more time working with the instructor in the classroom was very useful to me & \footnotesize 62.5\% \\
\\ \footnotesize In lecture-based model – I can always have a one-on-one interaction with the instructor after class or during instructor office-hours and I don't see any additional benefit of flipped-classroom over lecture-based model in terms of one-on-one interaction with the instructor & \footnotesize 37.5\% \\ \hline

\multicolumn{2}{|p{18cm}|}{\textbf{\small Q3: What is your experience with the flipped-classroom based teaching methodology in-terms of motivation, excitement and engagement level?}} \\ \hline
\footnotesize Flipped-classroom based teaching methodology and instruction model increased my motivation, excitement and overall engagement level & \footnotesize 45.0\% \\
\\ \footnotesize Flipped-classroom based teaching methodology and instruction model does not have any noticeable effect on my motivation, excitement and overall engagement level & \footnotesize 55.0\% \\ \hline

\end{tabular}
\end{center}
\end{table*}
The goals of our course is to teach and expose students to both technical and management issues encountered while developing a non-trivial (in the context of an academic setting and considering undergraduate students) software system for a real client and for a real-world problem. While we used lectures and textbooks to explain theory and fundamental concepts, the emphasis of our course was on practice and coach students on social aspects and practical know-how (create an experiential learning environment) through real-client project. Real client based project are common in Software Engineering courses (undergraduate and graduate level around the world) and is a proven instruction model \cite{Gnatz2003}\cite{Rosiene2006}\cite{Flener2006}\cite{Chen2011}\cite{Chase2007}\cite{Vanhanen2012}\cite{Hadfield2007}. However, there are some similarities and difference in the nature of projects in our course in comparison to other courses. 

We asked students to work on projects serving the need of the Institute. The goal was to identify problems and develop software-based solutions for problems encountered by the community (students, faculty and staff) at the Institute. The real customer for the students were the administrative departments (academic, finance and accounts, human resource, facility management) within the Institute. As shown in Table \ref{tab:realclient} (a survey response by students of the course on their views on real-world client project and deployment), about $68.5\%$ of the respondents mentioned that developing a solution aimed at serving the needs of the Institute increased their motivation substantially. Our aim was to create an environment (to the extent possible) in which the customer is not a researcher or practitioner of software engineering and is from functions such as finance, human resource and student affairs administration.   
Figure \ref{fig:realclientp} displays the list of projects developed by students of SE course and real users and customers with whom the students interacted. Students developed software such as TA allocation system, classroom and time-table scheduling, hostel allocation system, online assignment submission and grading tool, equipment purchase and leave management systems. 
\section{Studio-Based Learning}
Figure \ref{fig:framework} shows the seating arrangement (each group called as a studio sitting together in a circular arrangement) of the students in the class. The seating arrangement was different than the lecture-based model and is motivated by studio-based learning model applied in architecture schools. Nurkkala et al. present their experiences on teaching Software Engineering using studio-based learning (SBL) instruction model (Taylor University, Computer Science and Engineering curriculum) \cite{Nurkkala2011}. Their approach called as Software Studio is motivated by the need to train students as professional software engineers (producing professional quality software) in a class-room and university education setting \cite{Nurkkala2011}. Carter et al. present a review of studio-based learning in computer science and mention that SBL is becoming increasingly popular in computer science education \cite{Carter2011}. Table \ref{tab:studio} displays the results of survey conducted on studio-based learning aspect of the course. More than $70\%$ of survey respondents mention that the design critique and review sessions in studio-based learning method were quite in-depth and critical in comparison to presentation and review sessions in traditional (lecture-based) courses and helped in increasing my understanding of the concepts. More than $50\%$ of survey respondents mention that the studio-based teaching method and instruction model increased their motivation, excitement and overall engagement level. e believe that studio-based learning is a good match for a practice oriented and applied field like SE and the experience of studio-based learning can be enhanced by investing in infrastructure (different than traditional lecture-based model) facilitating studio-based working. However, we notice that studio-based learning in a large classroom increases instructor load and a larger teaching team is required to effectively conduct the classes. 
\section{Inverted or Flipped Classroom Model}
We believe that studio-based learning and real-client based projects in a large-team setting has synergy and complements with an inverted or flipped (also called as reverse instruction model) teaching methodology. Gannod et al. present their experiences with an inverted classroom model in the context of a software engineering curriculum at Miami University and mention that inverted classroom teaching environment mixes the use of technology with hands-on activities \cite{Gannod2008}. One of our motivations was to bring self-reading and self-learning (by watching video lectures and reading assigned text material) which is essential in today's world. Table \ref{tab:flipped} displays the result of survey conducted on inverted classroom model. We assigned video lectures and text material to students which are already present on YouTube or Internet (plenty of material is available on the topic as UG level SE course is a foundation level course in Computer Science programs) as we did not feel the need of creating our content. More than $60\%$ of the survey respondents mention that "since instructor is not lecturing, I have more opportunity of a one-on-one interaction with the Instructor in contrast to the lecture-based model. Having more time working with the instructor in the classroom was very useful to me". One of the challenges that we faced is that it is hard to know the students who are watching the video lectures. 
\bibliographystyle{plain}
\bibliography{ARXIV2013}
  
\end{document}